\ifx\pdfvariable\undefined
\pdfoutput=1
\fi
\AtBeginDocument{\paperdetails{
  submitted=2017-03-31,
  published=2017-08-07,
  year=2018,
  volume=2,
  issue=1,
  articlenumber=1,
}}
\documentclass[english]{programming}

\usepackage[backend=biber]{biblatex}
\usepackage{wrapfig}
\usepackage{multicol}
\lstdefinelanguage[programming]{TeX}[AlLaTeX]{TeX}{%
  deletetexcs={title,author,bibliography},%
  deletekeywords={tabular},
  morekeywords={abstract},%
  moretexcs={chapter},%
  moretexcs=[2]{title,author,subtitle,keywords,maketitle,titlerunning,authorinfo,affiliation,authorrunning,paperdetails,acks,email},
  moretexcs=[3]{addbibresource,printbibliography,bibliography},%
}%
\newcommand*{\CTAN}[1]{\href{http://ctan.org/tex-archive/#1}{\nolinkurl{CTAN:#1}}}

\usepackage{amsmath}
\usepackage{stmaryrd}
\usepackage{latexsym}
\newcommand\pfun{\mathrel{\ooalign{\hfil$\mapstochar\mkern5mu$\hfil\cr$\to$\cr}}}

\newcommand{\eg}{\emph{e.g.,}\xspace}
\newcommand{\ie}{\emph{i.e.,}\xspace}

\newcommand{\ct}[1]{{\textsf{#1}}\xspace}

\newcommand{\globalExtension}{\ct{global}}
\newcommand{\fn}[1]{\mathit{#1}}
\newcommand{\func}[3]{\fn{#1} : #2 \to #3}
\newcommand{\pfunc}[3]{\fn{#1} : #2 \pfun #3}
\newcommand{\stack}{\bar{m}}
\newcommand{\ext}{e}
\newcommand{\exts}{\bar{e}}
\newcommand{\concat}{\frown}

\newcommand{\Class}{\mathcal{C}}
\newcommand{\Signature}{\mathcal{S}}
\newcommand{\Method}{\mathcal{M}}
\newcommand{\CallStack}{\Gamma}
\newcommand{\Ext}{\mathcal{E}}

\newcommand{\superclassInv}{\fn{superclass}^{-1+}}
\newcommand{\superclasses}{\fn{superclass}^{+}}

\begin{document}

\title{Scoped Extension Methods in Dynamically-Typed Languages}

\author[a]{Guillermo Polito}
\authorinfo[figures/guille]{is research engineer at CNRS working currently in the RMoD~(\url{http://rmod.lille.inria.fr}) and Emeraude (\url{http://www.cristal.univ-lille.fr/emeraude/}) teams. His research targets programming language abstractions and tool support for modular long-lived systems. For this, he studies how reflective systems can evolve while maintaining these properties. He is interested in how these concepts combine with distribution and concurrency. Contact him at \email{guillermo.polito@univ-lille1.fr}.}
\affiliation{Univ. Lille, CNRS, Centrale Lille, Inria, UMR 9189 - CRIStAL - Centre de Recherche en Informatique Signal et Automatique de Lille}
\affiliation{Inria Lille - Nord Europe, UMR 9189 - CRIStAL - Centre de Recherche en Informatique Signal et Automatique de Lille}

\author[b]{St\'ephane Ducasse}
\authorinfo[figures/stef]{is directeur de recherche at Inria. He leads the RMoD~(\url{http://rmod.lille.inria.fr}) team. He is expert in two domains: object-oriented language design and reengineering.  He worked on traits, composable groups of methods. Traits have been introduced in  Pharo, Perl, PHP and  under a variant into Scala, Fortress of SUN Microsystems.  He is also expert on software quality, program understanding, program visualisations, reengineering and metamodeling. He is one of the developer of Moose, an open-source software analysis platform \url{http://www.moosetechnology.org/}.  He created \url{http://www.synectique.eu/} a company building dedicated tools for advanced software analysis. He is one of the leader of Pharo~(\url{http://www.pharo.project.org/}) a dynamic reflective object-oriented language supporting live programming. The objective of Pharo is to create an ecosystem where innovation and business bloom. He wrote several books such as Functional Programming in Scheme, Pharo by Example, Deep into Pharo, Object-oriented Reengineering Patterns, Dynamic web development with Seaside. Contact him at \email{stephane.ducasse@inria.fr}}
\author[c]{Luc Fabresse}
\authorinfo[figures/luc]{is associate professor in the CAR research theme (\url{http://car.mines-douai.fr}) at the Mines-Telecom Institute, Mines Douai, France. His researches aims at easing the development of mobile and constrained software using dynamic and reflective languages such as Pharo. One of his goal is to support live programming of mobile and autonomous robots in an efficient way. He is the co-author of multiple research papers (\url{http://car.mines-douai.fr/luc}) and he concretizes all these ideas (models and tools) in the PhaROS plateform (a Pharo client for the Robotics Operating System) to develop, debug, test, deploy, execute and benchmark robotics applications. Each year, Luc also gives computer science lectures, co-organizes events (technical days, conferences, ...) and promotes Smalltalk as an ESUG (European Smalltalk User Group) board member. Contact him at \email{luc.fabresse@mines-douai.fr}}
\affiliation{Mines Douai, Mines-Telecom Institute}
\author[b]{Camille Teruel}
\authorinfo[figures/camille]{is a young researcher and software engineer at Foretagsplatsen. His works reflection control for security. His main domain of work are on Meta-Object protocols for security and isolation. Contact him at \email{camille.teruel@gmail.com}}

\authorrunning{G. Polito, S. Ducasse, L. Fabresse, C. Teruel} 


\paperdetails{
  perspective=sciencetheoretical,
  area={General-purpose programming},
}



\maketitle
\pdfsuppressptexinfo=-1

\begin{abstract}
\textbf{Context.}
An extension method is a method declared in a package other than the package of its host class.
Thanks to extension methods, developers can adapt to their needs classes they do not own: adding methods to core classes is a typical use case. This is particularly useful for adapting software and therefore to increase reusability.\\
\textbf{Inquiry.}
In most dynamically-typed languages, extension methods are globally visible.
Because any developer can define extension methods for any class, naming conflicts ocur:
if two developers define an extension method with the same signature in the same class, only one extension method is visible and overwrites the other.
Similarly, if two developers each define an extension method with the same name in a class hierarchy, one overrides the other.
To avoid such ``accidental overrides'', some dynamically-typed languages limit the visibility of an extension method to a particular scope.
However, their semantics have not been fully described and compared.
In addition, these solutions typically rely on a dedicated and slow method lookup algorithm to resolve conflicts at runtime.\\
\textbf{Approach.}
In this article, we present a formalization of the underlying models of Ruby refinements, Groovy categories, Classboxes, and Method Shelters that are scoping extension method solutions in dynamically-typed languages.\\
\textbf{Knowledge.}
Our formal framework allows us to objectively compare and analyze the shortcomings of the studied solutions and other different approaches such as MultiJava.
In addition, language designers can use our formal framework to determine which mechanism has less risk of ``accidental overrides''.\\
\textbf{Grounding.}
Our comparison and analysis of existing solutions is grounded because it is based on denotational semantics formalizations.\\
\textbf{Importance.}
Extension methods are widely used in programming languages that support them, especially dynamically-typed languages such as Pharo, Ruby or Python.
However, without a carefully designed mechanism, this feature can cause insidious hidden bugs or can be voluntarily used to gain access to protected operations, violate encapsulation or break fundamental invariants.
\end{abstract}

\section{Introduction}
\label{sec:introduction}

Extension methods are a popular feature in dynamically-typed object-oriented languages.
An extension method is a method that a developer adds to a class which he does not own.
Variants of extension methods are available in many dynamically-typed languages: it is known as \emph{open classes} in Ruby~\cite{Mats01a}, \emph{categories} in Objective-C~\cite{Pins91a} and Groovy~\cite{Koen07a}, and \emph{extension methods} in Smalltalk~\cite{Gold84a} and Pharo~\cite{Blac09a}.
Other languages such as Golo~\cite{Pong13a}, which is a dynamically-typed language but offering only static method resolution, offers class extensions named class 'augmentations' but only supporting function addition.

Extension methods are globally visible in most existing implementations, causing two problems: \emph{accidental overwrites} and \emph{accidental overrides}.
An \emph{accidental overwrite} happens when two developers define an extension method with the same signature in the same class: in this case a conflict occurs and one method overwrites the other.
An \emph{accidental override} happens when two developers define an extension method with the same signature in the same class hierarchy: one method overrides the other. We call such overrides \textit{accidental} because they happen silently and unintentionally.
Another common problem is the absence of dependency declarations between extension methods and their callers.
Together with the global visibility of extension methods, this promotes the emergence of hidden dependencies that are difficult to track, especially in a dynamically-typed language.

One way to solve these problems is to assign each extension method a particular scope.
Variants of scoped extension methods have already been discussed in the literature with the Classbox model \cite{shortBerg03a,shortBerg05b,Berg05c}, the Method Shelters model \cite{Akai12a} and Matriona class extensions \cite{Spri16a}. In addition, scoped extension methods have been implemented in Ruby since version 2.1 and in Groovy.
These variants, however, have different semantics that must be well understood by the developers.
To the day of this writing, there is no clear description and comparison of their semantics as well as pros and cons of their impact on the way applications are built.
In addition, these variants rely on dedicated method lookup algorithms to resolve conflicts at runtime and tend to have a negative impact on speed.

In this article, we study the semantic differences between variants of scoped extension methods.
We scope our analysis to solutions in dynamically-typed languages.
We acknowledge solutions for this problem exist also in the context of statically-typed languages \cite{shortClif00a, shortWart06a, Duco07a}, but they are not directly applicable to dynamically-typed languages because they rely on static type information.
For the sake of completion, we finally compare solutions for both dynamically and statically typed languages.

The main contributions of this paper are:
\begin{itemize}
\item a precise study of the problems induced by extension methods (\autoref{sec:problem});
\item an in-depth study of existing solutions analyzing their main characteristics~(\autoref{sec:solutions});
\item a formalization and comparison of the underlying models of these solutions~(\autoref{sec:lookupFormalization} and \autoref{sec:analysis});
\item an example of how such formalisation can be used in the form of a metric to estimate the risk of accidental overrides~(\autoref{sec:minimizingAOS}).
\end{itemize}

\section{Extension Methods}
\label{sec:problem}

This section presents the extension method mechanism in detail.
First, we give some common use cases of extension methods.
Then, we show some problems induced by globally visible extension methods.

\subsection{Usage of Extension Methods} \label{sec:petitparser}

We show the advantages of extension methods with examples taken from \emph{PetitParser}, a parser combinator library for Pharo \cite{shortReng10c}.
In PetitParser, parsers are modeled as objects and parser combinators accept one or several parsers to produce a new composed parser.
Examples of combinators include ``\ct{,}'' to sequence two parsers and ``\ct{star}'' to repeat a parser zero or more times. For example, the following piece of code shows how we can create a parser that accepts the regular expressions of the form \textbf{ab*}\footnote{The syntax to denote a character in Smalltalk is the character itself preceded by a dollar sign (\$).}:

\begin{lstlisting}
(PPLiteralObjectParser on: $a), (PPLiteralObjectParser on: $b) star.
\end{lstlisting}

\begin{figure}[htb]
  \begin{center}
    \includegraphics[width=9cm]{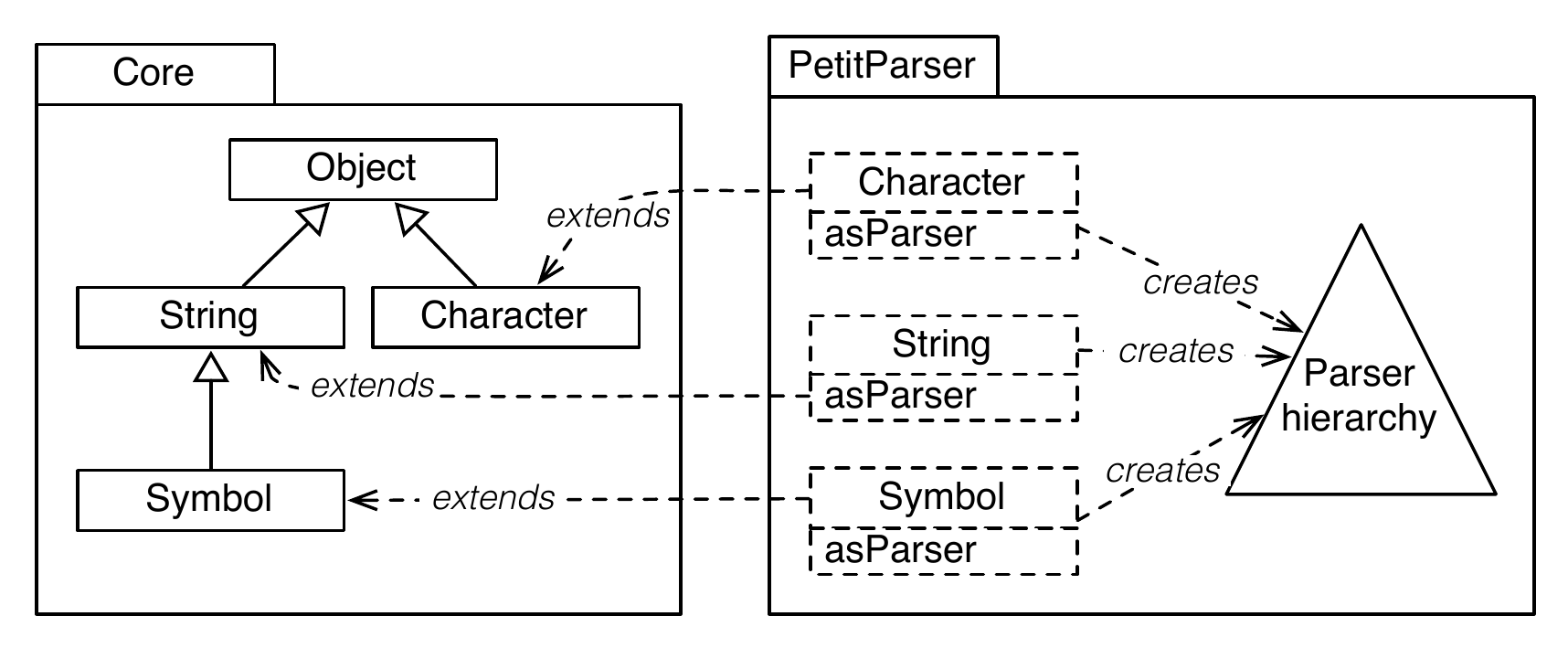}
    \caption{The PetitParser parser combinator library defines \ct{asParser} extension methods on core classes to create various kinds of parsers.}
    \label{fig:ppClassDiagram}
    \vspace{-0.9cm}
  \end{center}
\end{figure}

\paragraph{As syntactic sugar.} In addition to these combinators, \emph{PetitParser} defines convenient \ct{asParser} extension methods to some core classes.
These extension methods create parsers depending on the receiver (see \autoref{fig:ppClassDiagram}).
For example, the \ct{asParser} extension method defined in the class \ct{Character} returns\footnote{The character \lstinline|^| stands for return in Smalltalk syntax.} a parser that accepts the receiver character.

\begin{lstlisting}
Character >> asParser
    ^ PPLiteralObjectParser on: self
\end{lstlisting}

Together, combinators and \ct{asParser} extension methods give a readable DSL-like syntax.
For example, the following expression creates the same parser as before for \textbf{ab*}:

\begin{lstlisting}
$a asParser, $b asParser star
\end{lstlisting}

In this example, extension methods \ct{asParser} act as syntactic sugar \ie~\ct{\$a asParser} has the same meaning as \ct{PPLiteralObjectParser on: \$a}.

\paragraph{To improve extensibility.}
Extension methods can also improve code quality by making classes polymorphic together.
Consider the following code:

\lstset{columns=fullflexible,tabsize=3,showtabs=false}
\begin{lstlisting}
MyParser>>one: a thenMany: b
  ^ a asParser , b asParser star.

MyParser>>id																																							MyParser>>int
   ^ self 																																																				^ self
        one: #uppercase												                                  																								one: ($1 to: $9)
        thenMany: #letter																                             																		thenMany: #digit
\end{lstlisting}

In the \ct{MyParser} class, the \ct{one:thenMany:} method takes as parameter two objects that can be converted into parsers and returns a new parser.
The \ct{id} and \ct{int} methods use that first method to build custom parsers.
The method \ct{id} sends the message \ct{one:thenMany:} with two symbols (\ct{uppercase} and \ct{letter}) while the method \ct{int} sends the same message with an interval and a symbol.
Extension methods are useful here as they allow any developer to add the method \ct{asParser} to any class and pass instances of this class to \ct{one:thenMany:}.

\paragraph{To adapt classes interface.}
The Adapter pattern adapts the interface of an existing class to work with other classes without modifying its source code.
The classic realization consists in creating an adaptor class whose instances are used to wrap the instances of the adapted class whenever needed at the expenses of obtaining a different identity.
Extension methods permit a class interface to be adapted without relying on an adapter class.
Instances of the adapted class can be used directly as they do not need to be wrapped with an adapter object. Therefore, object identity is preserved and less objects are created (no adapters).


\paragraph{Monkey-patching.}
If a third-party library or framework has a bug, a developer can create overwriting extension methods to correct the bug.
This technique is known as \emph{monkey patching}. While monkey patching is often recognized as a bad practice in developer communities, it is occasionally useful.

\subsection{Problems of Globally Visible Extension Methods}
\label{sec:global-visibility}

Despite all the benefits that extension methods bring to developers, they can also cause several conflicts and headaches, specially when their usage is not controlled or scoped. Most implementations of extension methods, such as the ones present in various Smalltalk dialects, Ruby (before the introduction of \emph{refinements}) and Objective-C, make extension methods globally visible. This can lead to undeclared dependencies, \emph{accidental overwrites} and \emph{accidental overrides}.

\paragraph{Undeclared dependencies.}
Once an extension method is loaded it is globally visible. The method can be called from any class of any loaded package without any form of declaration.
This means that an application can work correctly in the developer's environment and fail once deployed because the application depends on an extension method from a non-loaded package.
The absence of declaration favors the emergence of such hidden dependencies.

\begin{figure}[htb]
\centering
\includegraphics[width=0.60\linewidth]{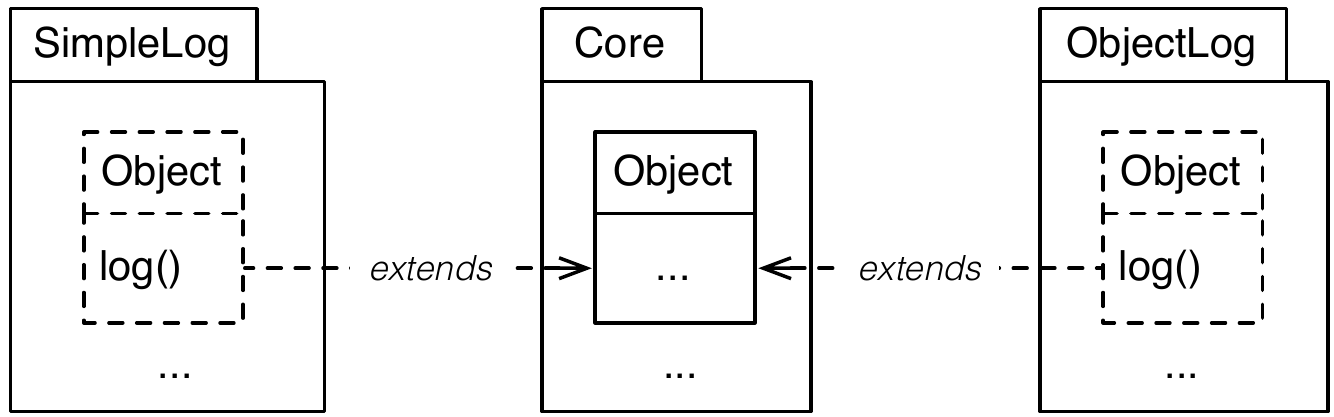}
\caption{An example of accidental overwrite. Two packages each declare an extension method \ct{log} for the class \ct{Object}.}
\label{fig:overwrite}
\vspace{-0.8cm}
\end{figure}

\paragraph{Accidental overwrites.}
Extension methods defined by different packages may conflict in two different ways.
The first kind of conflict arises when two packages each define an extension method with the same signature in the same class.
In this case, one extension method replaces the other.
We call this situation an \emph{accidental overwrite}.
We show an example in \autoref{fig:overwrite}.
The class \ct{Object} is part of the package \ct{Core}.
A package \ct{SimpleLog} declares an extension method \ct{log} for the class \ct{Object}.
This package is a logging framework that records the string representation of an object in a log file.
The package \ct{ObjectLog} declares another extension method \ct{log} for the class \ct{Object}.
This latter package is another logging framework that serializes objects in a log file for later analysis.
Both extension methods conflict and the two logging frameworks cannot be loaded at the same time.

Even though these name clashes happen sparingly, they are difficult to anticipate, especially when considering package co-evolution in large projects involving several packages.

\begin{figure}[htb]
\begin{center}
\includegraphics[width=0.7\linewidth]{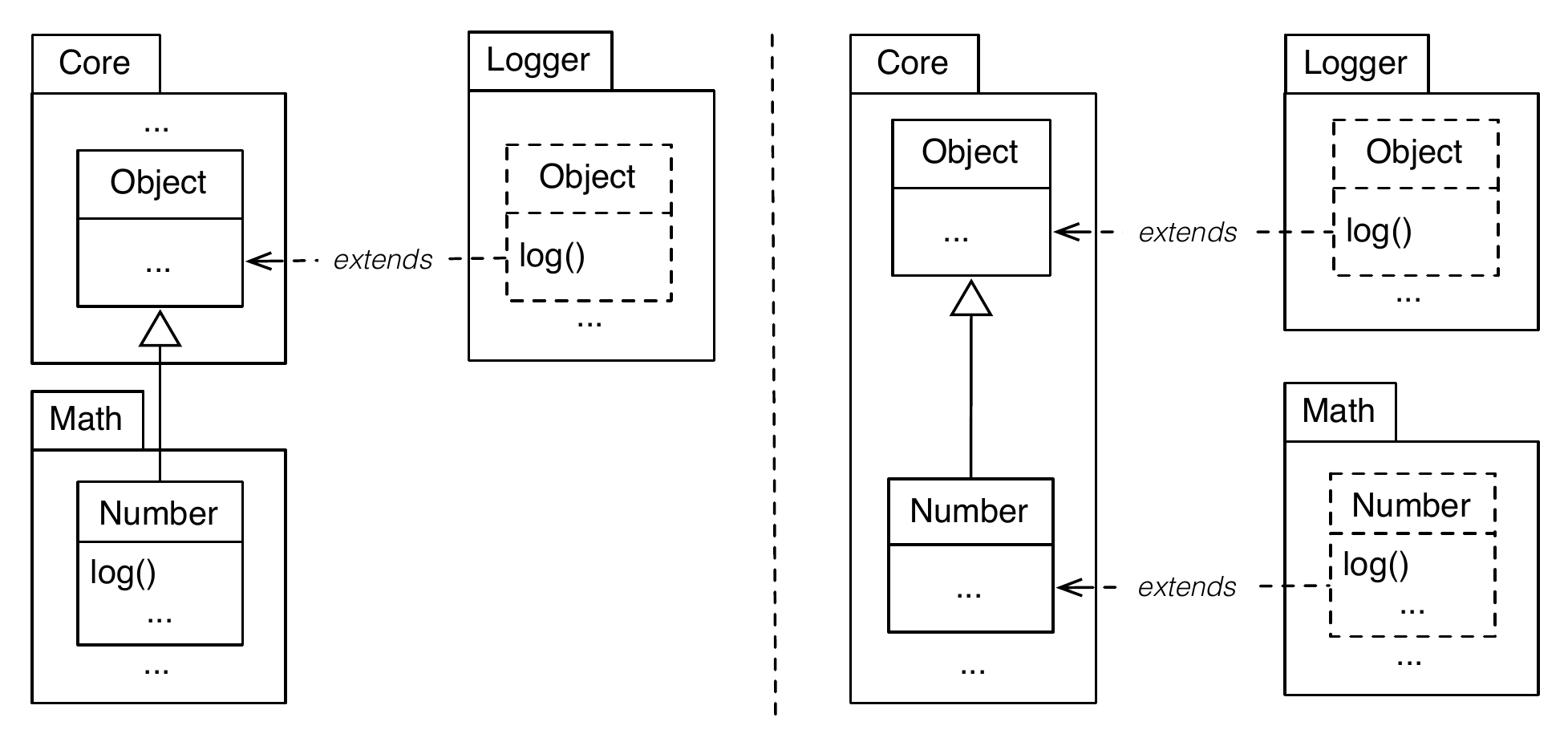}
\caption{Two examples of accidental overrides.
To the left, a regular method accidentally overrides an extension method.
To the right, an extension method accidentally override another extension method.}
\label{fig:override}
\vspace{-0.8cm}
\end{center}
\end{figure}

\paragraph{Accidental overrides.}
The second kind of conflict arises when an extension method overrides another extension method defined higher in the class hierarchy.
We depict two examples of such overrides in \autoref{fig:override}.
On the left part, a regular method \ct{log} in package \ct{Math} accidentally overrides an extension method in its superclass declared in package \ct{Logger}.
While \ct{Logger}'s extension method \ct{log} prints the receiver object in some log file, \ct{Math}'s extension method \ct{log} computes the logarithm of a number.
When they send the message \ct{log} to an object, users of the \ct{Logger} package expects that the extension method of \ct{Logger} is taken into account.
However, \ct{Number} class, as a subclass of \ct{Object} overrides that \ct{log} method in package \ct{Math}.
On the right part of \autoref{fig:override}, an extension method in package \ct{Math} overrides another extension methods in package \ct{Logger}. The situation is very similar to the previous one.
The package \ct{Math} and \ct{Logger} are unaware of each other so none of them know that \ct{Math}'s extension method overrides \ct{Logger}'s.

\medskip

Large programs usually involve multiple concerns and domains, each coming with its own terminology.
Accidental overwrites and overrides happen when these terminologies overlap.
In the context of extension methods, the probability of accidental overwrites and overrides is large because any package can declare an extension method for any class.
Accidental overrides are a form of interference between packages which is more insidious than accidental overwrites.
Indeed, an accidental overwrite is easily noticeable because the client packages are likely to break upon the first invocation of the overwriting method.
Accidental overrides are much less noticeable because they affect only instances of the class defining the overriding method.
Note that this problem only appears because defining a method implies overriding methods with the same signature upper in the hierarchy.

Since multiple parties can enhance the interface of any class, one party should not be able to override the methods defined by an unrelated party it is not aware of.
In other words, extension methods need to be scoped.

\section{Existing Mechanisms for Scoped Extension Methods}
\label{sec:solutions}

Because extension methods with global visibility exhibit the above-mentioned problems, several implementations propose a narrower visibility.
This section describes five of these solutions we selected because they are representative of five different scoping strategies.
Depending on the solutions, the scope of activation of extension methods is either lexical or dynamic.
In solutions where the scope of activation is lexical, the set of extension methods that are active at a given point is determined statically.
In solutions with a dynamic scope of activation, the set of extension methods active at a given point depends on a dynamic context: the call stack.
Dynamic scoping is necessary to support a property called \emph{local rebinding}.

First we present the local rebinding property and some of its weaknesses.
Then, we show three solutions that expose the local rebinding property: \emph{Classboxes}~\cite{shortBerg03a,shortBerg05b}, \emph{Method Shelters}~\cite{Akai12a} and Groovy's \emph{categories}~\cite{Koen07a}.
Finally, we present Ruby's \emph{refinements} and Selector Namespaces~\cite{Wirf96a} where extension method activation is determined lexically.

\subsection{Definitions}
\noindent{}
In the following, we use the following terms:
\begin{description}
	\item[\textit{Package.}] We call package, the language-specific unit of deployment that gathers definitions of classes and other constructs from the language. Different packages are potentially maintained by different parties. A package also declares dependencies to other packages by importing some definitions.
	\item[\textit{Class extension.}] A class extension is a named set of extension methods that apply to the same class. We do not consider addition of instance variables.
	\item[\textit{Extension group.}] An extension group is a named set of extension methods that may apply to different classes.
\end{description}

\subsection{Introduction to Local Rebinding}
\begin{figure}[htb]
\centering
\includegraphics[width=0.8\linewidth]{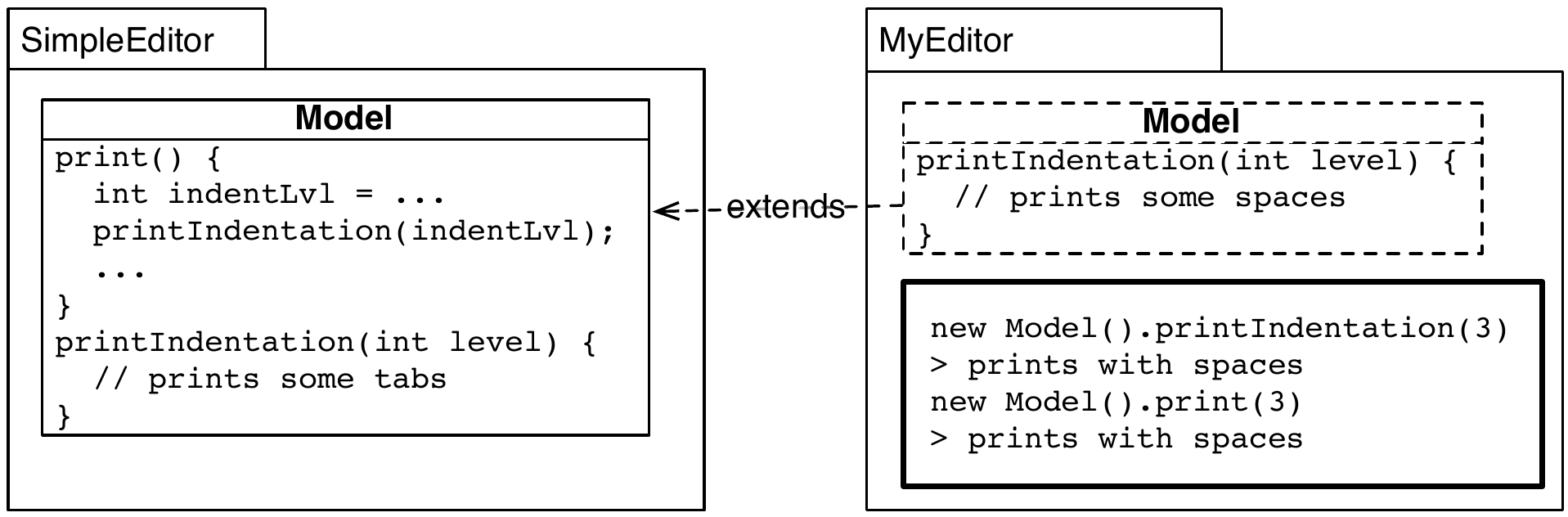}
\caption{With local rebinding, changes made by an extension method are applied in case of indirect calls.}
\label{fig:localRebindingExemple}
\vspace{-0.2cm}
\end{figure}

\emph{Local rebinding} is a method-lookup algorithm first defined in the Classbox model~\cite{shortBerg03a,shortBerg05b} and refined in the Method Shelters model~\cite{Akai12a} and hierarchical layer-based class extensions \cite{Spri16a}. This property permits extension methods to override regular methods in a contextual manner.
An active extension method takes precedence over regular methods, even for indirect calls.
In \autoref{fig:localRebindingExemple}, the \ct{MyEditor} package defines an extension method \ct{printIndentation(int)} that redefines the one in the original package.
This extension method prints spaces instead of tabs.
When invoked from within this package, this redefinition is taken into account, even in indirect calls:
when invoking the \ct{print()} method defined in the \ct{SimpleEditor}  package, the redefined version of \ct{printIndentation(int)} will be executed and not the one defined in the \ct{SimpleEditor} package.

With local rebinding, the lookup algorithm may have to dispatch to different methods in different contexts.
In technical terms, when the signature of an extension method \emph{e} matches the one of a method \emph{m} of the extended class, local rebinding ensures that \emph{e} overrides \emph{m} during the dynamic extent of message sent by importers of \emph{e}.
The method lookup algorithm has to access this contextual information to determine the active extension methods.
Such a method lookup algorithm can be implemented either by inspecting the call stack or by storing the set of active extension methods in a dynamic variable.

\subsection{Illustrating local rebinding stack behavior}

 \begin{figure}[htb]
    \centering
    \includegraphics[width=0.7\linewidth]{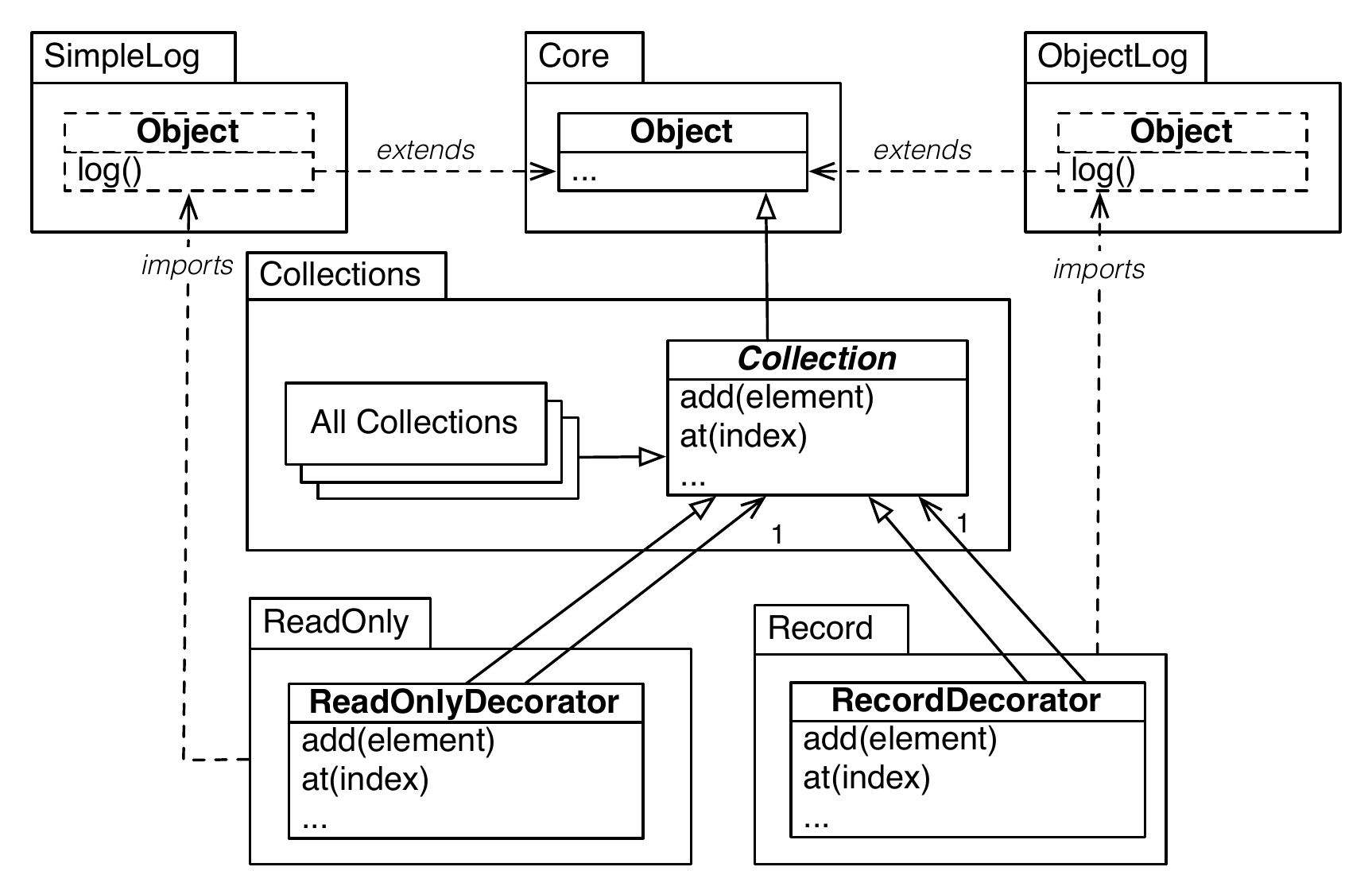}
    \caption{Decorating collections: two potentially conflicting extensions selected by on stack state.}
    \label{fig:collections}
    \end{figure}

Consider the example depicted in \autoref{fig:collections}.
A \ct{Collections} package defines common collections and an abstract class \ct{Collection}.
Two packages \ct{ReadOnly} and \ct{Record} each define a collection decorator.

The read-only decorator disables all operations that mutate the decorated collection.
When one of these operations is invoked, the read-only decorator logs the attempt using the logging facility of the \ct{SimpleLog} package and throws an error.
The record decorator just logs the operations done on the decorated collection using the logging facility of the \ct{ObjectLog} package for latter analysis as shown below in pseudo-code.

\lstset{columns=fullflexible,tabsize=3,showtabs=false}
\begin{lstlisting}
ReadOnlyDecorator>>at(index)																																	ReadOnlyDecorator>>add(element)
    return decoree.at(index)																																										'adding failed'.log();
																		throw IllegalWrite()

RecordDecorator>>at(index)																																					RecordDecorator>>add(anObject)
    { 'accessing'. decoree. index }.log();																									{'adding'. decoree. anObject }.log();
    return decoree.at(index)																																											return decoree.add(index)
\end{lstlisting}

If a client application uses both decorators together, one \ct{log()} method is likely to contextually override the other.
This is the case when one decorator decorates the other.
In this case, the composition order matters because it impacts the call stack and thus the extension methods that are active when looking up \ct{log()}.

\vspace{5mm}
\begin{tabular}{cl}
             			& \ct{list = new List([1,2,3,4]);}									\\
\emph{Case 1}     	& \ct{(new ReadOnlyDecorator(new RecordDecorator(list))).at(3);}		\\
\emph{Case 2}     	& \ct{(new RecordDecorator(new ReadOnlyDecorator(list))).add(5);}	\\
\end{tabular}

\vspace{5mm}

\begin{tabular}{ll}
\emph{Stack for Case 1}		& \emph{Stack for Case 2}			\\
~~~2.~ \ct{RecordDecorator.at()}		& ~~~2.~ \ct{ReadOnlyDecorator.add()}	\\
~~~1.~ \ct{ReadOnlyDecorator.at()}		& ~~~1.~ \ct{RecordDecorator.add()}
\end{tabular}

\vspace{5mm}

In Case 1, a read-only decorator decorates a record decorator that decorates a list.
When sending the \ct{at(3)} message to the read-only decorator, first its \ct{at()} method transfers the request to the record decorator.
The \ct{at()} method of the record decorator then tries to log this operation.
At this point, two method activations are at the top of the call-stack:
first an activation of the \ct{at()} method of record decorator, then an activation of the \ct{at()} method of the read-only decorator.
Since each package defining the \ct{at()} method imports a different \ct{log()} extension method, the lookup algorithm must decide which one to select.
A similar situation occurs with Case 2 with another call order.

We now study two strategies to select a method in case of ambiguities: \emph{bottom-up local rebinding} and \emph{top-down local rebinding}.

\subsection{Bottom-up Local Rebinding}

The first strategy gives precedence to extension methods imported by callers (\ie appearing first in the call stack).
We refer to this strategy as \emph{bottom-up local rebinding}.
This is the strategy of the Classbox~\cite{shortBerg03a,shortBerg05b} and Method Shelters \cite{Akai12a} models.
In the context of \autoref{fig:collections}, this means that the \ct{log()} method of the \ct{SimpleLog} package is selected in Case 1 and the \ct{log()} extension method of the \ct{ObjectLog} package is selected in Case 2.
This strategy implies that \emph{client code may override other extension methods defined in any package}.
As the developer of a package, your methods can be overridden by a package that is indirectly using yours.
Consequently, this forces a developer to know the implementation of all the packages it uses (even indirectly) to prevent himself from creating accidental contextual overrides.
This raises a tension with information hiding at the package-level and precludes local reasoning.

\subsubsection{Classboxes}

A classbox is a modular construct defining classes and class extensions, taking the role of a package.
A classbox can define at most one \emph{class extension} per imported class.
This prevents useful ways to group related extension methods (See \autoref{sec:dependencies}).
A classbox can import class extensions from other classboxes.
Classboxes have been devised to facilitate handling of unanticipated changes \ie a client classbox pushes modifications to other classboxes.
Used sparingly, classboxes allows developers to customize the implementation of external packages.
However, if used extensively, accidental contextual overrides are likely to occur.

\subsubsection{Method Shelters}

The Method Shelters model \cite{Akai12a} builds upon the Classbox model by adding the ability to protect some extension methods from accidental contextual overriding.
A method shelter is a package that contains an \emph{exposed chamber} and a \emph{hidden chamber}.
Each chamber declares classes and methods, and they can import other method shelters.
Importing a method shelter brings the extension methods of its exposed chamber into the importing chamber.
Thus, only methods imported or declared in the exposed chamber can be contextually overridden by other method shelters.

We illustrate the behaviour of hidden chambers in \autoref{lst:shelterExample}.
In the figure, two definitions of division (\ct{/}) over integers coexist without accidental contextual override.
The default \ct{/} method of the \ct{FixNum} class defines euclidian division.
The \ct{Math} shelter redefines \ct{/} as exact division: the method returns a rational number.
The \ct{average} shelter imports the \ct{math} shelter in its hidden chamber.
The \ct{avg} method of \ct{Array} uses the exact division of the \ct{Math} shelter to compute the average of an array of integers.
Finally a \ct{client} shelter imports the \ct{average} shelter and computes the average of an integer array:
the computation results in a rational number.
The \ct{client} shelter is oblivious of the fact that the \ct{average} shelter uses the \ct{Math} shelter.
From its point of view, \ct{/} still refers to the standard euclidian division.

\lstset{columns=fullflexible,tabsize=3,showtabs=false}
\begin{figure}
\begin{lstlisting}
shelter :MathShelter																																																					shelter :ClientShelter do
  class Fixnum #fixed size integer in Ruby																				import :AverageShelter
    def /(x)																																																																							def calc
      Rational(self,x)																																																									p([1,2,3,4,5,6,7,8,9,10].avg) #prints "(11/2)"
        end																																																																									p(55/10) #prints 5
    end																																																																														end
end																																																																																		end

shelter :AverageShelter do																																											shelter_eval :ClientShelter do
  class Array																																																																									calc
    def avg																																																																						end
      sum = self.inject(0){|r,i| r + i}
      sum / self.size
    end
  end
  hide
  import :MathShelter
end
\end{lstlisting}
\caption{Example taken from \cite{Akai12a}.
Method shelters provide the ability to control which method can be overridden: extension methods declared after \ct{hide} cannot be overridden by client shelters}
\label{lst:shelterExample}
\end{figure}

Method shelters work as classboxes if a program uses only the exposed chambers, and thus, this means that the same problems arise.
On the other hand, putting all methods inside the hidden chamber prevents the redefinition of methods, avoiding the local rebinding property.

\subsection{Top-Down Local Rebinding}

The second strategy gives priority to extension methods imported by callees.
With this priority strategy, an extension method can be overridden in a called method.
In the context of \autoref{fig:collections}, this means that the \ct{log()} method of the \ct{ObjectLog} package is selected in Case 1 and the \ct{log()}  extension method of the \ct{SimpleLog} package is selected in Case 2.
We refer to this strategy as \emph{top-down local rebinding}.
This is the strategy of \emph{Categories} in Groovy~\cite{Koen07a}.

\subsubsection{Groovy Categories}

Groovy developers can define scoped extension methods in \emph{categories}.
A category defines a named extension that can be put into the scope of a block of code using the \ct{use} keyword.
When a \ct{use} block is entered, the category is activated by pushing it onto a thread-local stack variable.
This extension is popped from the thread-local stack of active extensions when the block is exited.
Upon method lookup, a method redefined in a category takes priority over the original method in the extended class.
In case of conflict between two extension methods in two categories, the method defined in the lastly-activated category (the one that is nearest to the top of the stack) is selected.
Moreover, a \ct{use} block can activate several categories. If there is conflicting methods in these categories, the first definition hides the others.

\bigskip

\subsection{Lexical extension activation}

This section presents scoped extension mechanisms using a lexical scope of activation, in contrast to the already presented models providing local-rebinding.
In these solutions, only extensions defined and imported explicitly in the current lexical scope are active during the execution of a program.
This kind of scoped extension methods is provided by \emph{refinements} in Ruby, and \emph{selector namespaces} in \emph{SmallScript}.
In the rest of this section we describe ruby refinements and selector namespaces as significant examples of these solutions.

\subsubsection{Subsystems and Selector Namespaces}

We report on the Subsystem proposal since it is probably the source of inspiration for SmallScript~\cite{Wirf96a}.
In the subsystem proposal \cite{Wirf96a}, method signatures (selectors) are decomposed in two parts: a \emph{value} and a \emph{name}.
A selector \emph{value} is the key that is used to actually identify methods.
A message send uses a selector \emph{value} to select a method with the same selector \emph{value}. This value is not known to the programmer.
A selector \emph{name} is the actual symbolic name used in Smalltalk code to refer to a selector \emph{value}.

A message send using a selector name dispatches to whichever selector value that is bound to it in the lexical scope.
In other words, message name resolution is static. Selectors are organized into hierarchies that support redefinition and shadowing.
When the Smalltalk compiler processes a message, it looks up the selector names in the current scope and any of its enclosing scopes and uses the selector value that is found.

Unfortunately, the lack of a more clear documentation for \emph{selector namespaces} prevents us from analyzing its properties more in detail.

\subsubsection{Ruby Refinements}

Since its first versions, Ruby supports globally visible extension methods under the name of \emph{open classes}.
Ruby 2.1 introduces scoped class extensions under the name of \emph{refinements}.
In refinements, only modules and classes importing a refinement can call its extension methods.
In addition, if a class imports a refinement in its body, this refinement is also active in the scope of the subclasses, even when the subclasses are defined in another package.
This propagation of visibility provides some common facilities to subclasses, a feature that may be useful in frameworks where an abstract class of the framework is subclassed by users.
Also, developers who subclass an external class should be aware of the refinements that are active in that class.
Surprisingly, while the sequence of active refinements can be determined statically, the implementation of refinements does the resolution dynamically with a dedicated and slower method lookup.
This choice may be due to other implementation constraints.

\section{Method Lookup Formalization of Scoped Extension Solutions}
\label{sec:lookupFormalization}

We presented in \autoref{sec:solutions} several models of scoped extension methods.
To study the different design choices of each model, we present in this section a formal specification of a method lookup algorithm for scoped extension methods.
Using this specification, we formalized the different strategies of the studied solutions to select active extensions and then select the method to execute.

\subsection{Notations and Base Model}
We use five semantic domains and three functions to model language entities and their relations. We use $\pfun$ to denote partial functions.
\[
{
  \begin{array}{l@{~~~}l}
    c \in \Class                      & \pfunc{superclass}{\Class}{\Class}                               \\
    s \in \Signature                  &  \pfunc{method}{ \Class \times \Signature \times \Ext}{\Method} \\
    m \in \Method              & \func{imports}{\Method}{\Ext^*}                                       \\
    \ext \in	\Ext                & \\
    \stack \in \CallStack \text{ where } \CallStack = \Method^* & \\
  \end{array}
}
\]

Classes are elements of $\Class$.
Signatures are elements of $\Signature$.
In a dynamically-typed language a signature usually consists of a name and a number of parameters.
Methods are modeled as elements of $\Method$ and extension groups are modeled as elements of $\Ext$.

The partial function $\fn{superclass}$ denotes the class-superclass relationship.
Because a class cannot inherit from itself, this function is acyclic.
For a class $c$ that has no superclass, $\fn{superclass}(c)$ is undefined: usually, only one such class exists in programming languages.

The partial function $\fn{method}$ denotes the existence of a method in a given context.
This function returns a method if a given extension defines such method with a given signature for a given class.
It is undefined if the extension defines no such method.

The function $\fn{imports}$ returns a sequence that models which extension imports are effective in the context of a method, in order of decreasing priority (we note $\mathcal{X}^*$ the set of finite sequences of elements of $\mathcal{X}$).
These imports could be declared for a single method, for a whole class, for a whole class hierarchy, for a whole package, \emph{etc}.
What matters is which one affects a method and this is what $\fn{imports}$ indicates.

Finally, call stacks are elements of $\CallStack$.
For the purpose of modelling lexical and local-rebinding mechanisms, knowing the method of a stack frame is enough so call stacks are modeled as a finite sequence of methods ($\CallStack = \Method^*$).
The first element of such a method sequence corresponds to the bottom of the call stack and the last element corresponds to the method that sent the message being looked up.
We use the notation $\bar{x}$ for sequences (\ie $\bar{x}=<\bar{x}_1,\dots,\bar{x}_{|\bar{x}|}>$).

\paragraph{Standard Lookup.}
The standard lookup algorithm for class-based dynamically-typed languages with single dispatch depends on the class of the receiver object and a method signature.
To take the dynamic scoping of classboxes and method shelters into account, a lookup algorithm must also consider the call stack.
Consequently, we model the lookup as:
\[ \pfunc{lookup}{\Class \times \Signature \times \CallStack}{\Method} \]

A method lookup fails whenever the $\fn{lookup}$ function is undefined.
To better distinguish between the different kinds of lookup algorithms, we divide the lookup in two steps.
The first step determines the sequence of active method extensions from a call stack.
The second step selects a suitable method to be executed among a sequence of extensions.
The $\fn{lookup}$ function is then defined using two functions: $\fn{activeExts}$ and $\fn{select}$ that represent these two steps.
\[
    \fn{lookup}(c,s,\stack) = \fn{select}(c,s,\fn{activeExts}(\stack)) \quad \text{where:}
    \begin{cases}
    \func{activeExts}{\CallStack}{\Ext^*}\\
    \pfunc{select}{\Class \times \Signature \times \Ext^*}{\Method}
    \end{cases}
\]

We can now describe different versions of the $\fn{activeExts}$ and  $\fn{selection}$ functions separately.
We call the different versions of $\fn{activeExts}$: \emph{active extensions strategies}; and the different versions of $\fn{select}$: \emph{method selection strategies}.

\subsection{Active Extensions Strategies}

We now review the different active extension strategies.
In the context of local rebinding, the lookup has to consider the chain of callers to find if one imports an extension with an overriding extension method.
The extension activation is dynamically-scoped. This means that the lookup algorithm traverses the call stack or uses a thread-local variable to determine active extensions.
The call-stack can be traversed bottom-up giving priority to callers imports, or top-down, giving priority to callees imports.
Without local rebinding, the extension activation is said to be lexical.
For each strategy, we consider that a global extension named $\globalExtension$ contains all regular methods and it is implicitly imported by default.

\begin{figure}[htb]
\begin{center}
\includegraphics[width=0.85\linewidth]{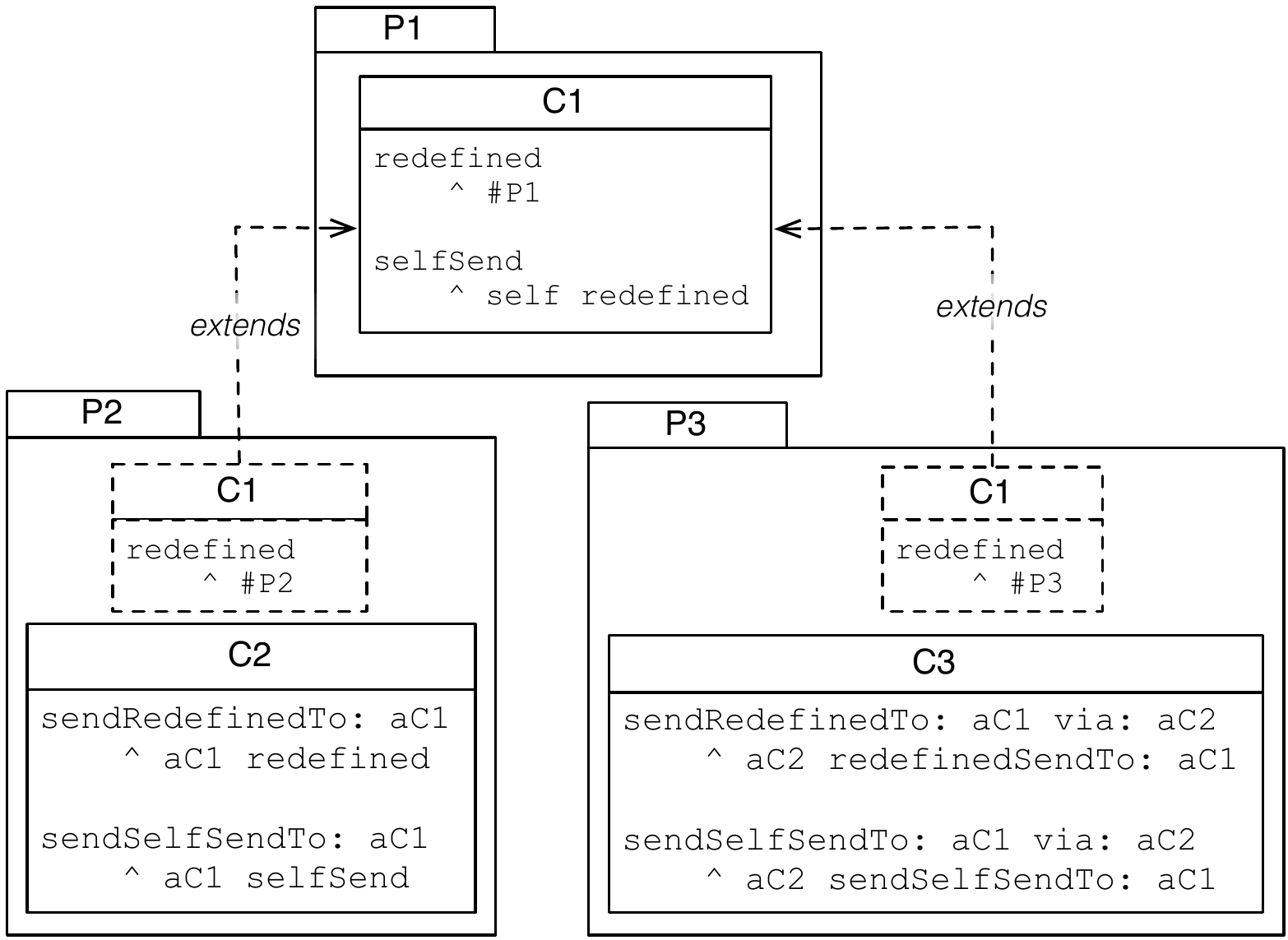}

\vspace{0.5cm}

\begin{tabular}{lccc}
& LR$\uparrow$ & LR$\downarrow$ & Lex\\
\hline
aC2 sendRedefinedTo: aC1 & \#P2 & \#P2 & \#P2\\
aC2 sendSelfSendTo: aC1 & \#P2 & \#P2 & \#P1\\
aC3 sendRedefinedTo: aC1 via: aC2 & \#P3 & \#P2 & \#P2\\
aC3 sendSelfSendTo: aC1 via: aC2 & \#P3 & \#P2 & \#P1\\
\end{tabular}
\caption{Active Extension Strategies by Example. This figure shows what are the results of four different expressions in the presence of the different active extension strategies.}
\label{fig:illustrating_activation_strategies}
\end{center}
\end{figure}

Besides the formalization, \autoref{fig:illustrating_activation_strategies} summarizes and illustrates active extension strategies through an example.
In the example, two packages \ct{P2} and \ct{P3} extend class \ct{C1} from package \ct{P1} with an override.
The table illustrates how four different method invocations behave in this scenario:
\emph{(a)} a class in \ct{P2} calls a redefined method of \ct{C1};
\emph{(b)} a class in \ct{P2} calls a method of \ct{P1} calling a redefined method of \ct{C1};
\emph{(c)} a class in \ct{P3} calls (a);
\emph{(d)} a class in \ct{P3} calls (b).
At runtime, this generates overrides between the redefined method in \ct{P2} and \ct{P3}.
We see that lexical activations use the definition available in the current lexical scope.
Local-rebinding, on the other side, will depend on the order of message sends in the stack.
Cases (c) and (d) give precedence to the method defined in \ct{P2} or \ct{P3}, depending on the local-rebinding strategy.

\paragraph{Bottom-up local rebinding.}
We first consider the extension activation strategy of bottom-up local rebinding as exemplified by Classboxes and also by Method Shelters exposed chambers.
The selection of active extensions for method shelters is more refined as it stops searching if one of the shelters is imported in a hidden chamber.
Here is the definition of the $\fn{activeExts_{lr\uparrow}}$ that computes the active extensions following this strategy:
\[
  \fn{activeExts_{lr\uparrow}}(\stack) = \fn{imports}(\stack_{1}) \concat \dots \concat \fn{imports}(\stack_{|\stack|}) \concat <\globalExtension>
\]

If the call stack is empty, that is if this is the first lookup of the associated thread, the function just returns the implicitly imported $\globalExtension$ extensions.
Otherwise, the function recursively concatenates the imports of each method in $\stack$ from the oldest method activation ($\stack_{1}$) to the newest ($\stack_{|\stack|}$).
Concatenation of sequence is noted ``$\concat$".
As a result of this bottom-up approach, extensions imported in the methods of the oldest stack frames come first.

\paragraph{Top-down local rebinding.}
Now, we consider top-down call-stack traversal as exemplified by Groovy categories.
Here is the definition of the $\fn{activeExts_{lr\downarrow}}$ that computes the active extensions following this strategy:
\[
  \fn{activeExts_{lr\downarrow}}(\stack) = \fn{imports}(\stack_{|\stack|}) \concat \dots \concat \fn{imports}(\stack_{1}) \concat <\globalExtension>
\]

Like with the function $\fn{activeExts_{lr\downarrow}}$, if call stack is empty $\fn{activeExts_{lr\downarrow}}$ returns the implicitly imported $\globalExtension$ extension.
Otherwise, the function recursively concatenates the imports of each method in $\stack$ from the newest method activation to the oldest.
As a result of this top-down approach, extensions imported in the methods of the newest stack frames come first.

\paragraph{Lexical extension activation.}
We finally consider the lexical extension activation strategy as exemplified by Ruby refinements.
A lexical extension activation means that the call-site is enough to know the active extensions.
It also means that the sequence of active extension is known statically.
The active extensions are the ones imported by the calling method, that is the first element of the sequence $\stack$.
\[
  \fn{activeExts_{lex}}(\stack) = \fn{imports}(\stack_1) \concat <\globalExtension>
\]

Choosing one of these three active extensions determination strategies  (bottom-up local rebinding, top-down local rebinding, lexical) determines which method extensions are active during a message send.
The next step of the lookup is to choose a method among these extensions.

\subsection{Method Selection Strategy}

\begin{figure}[htb]
    \centering
    \includegraphics[width=0.55\linewidth]{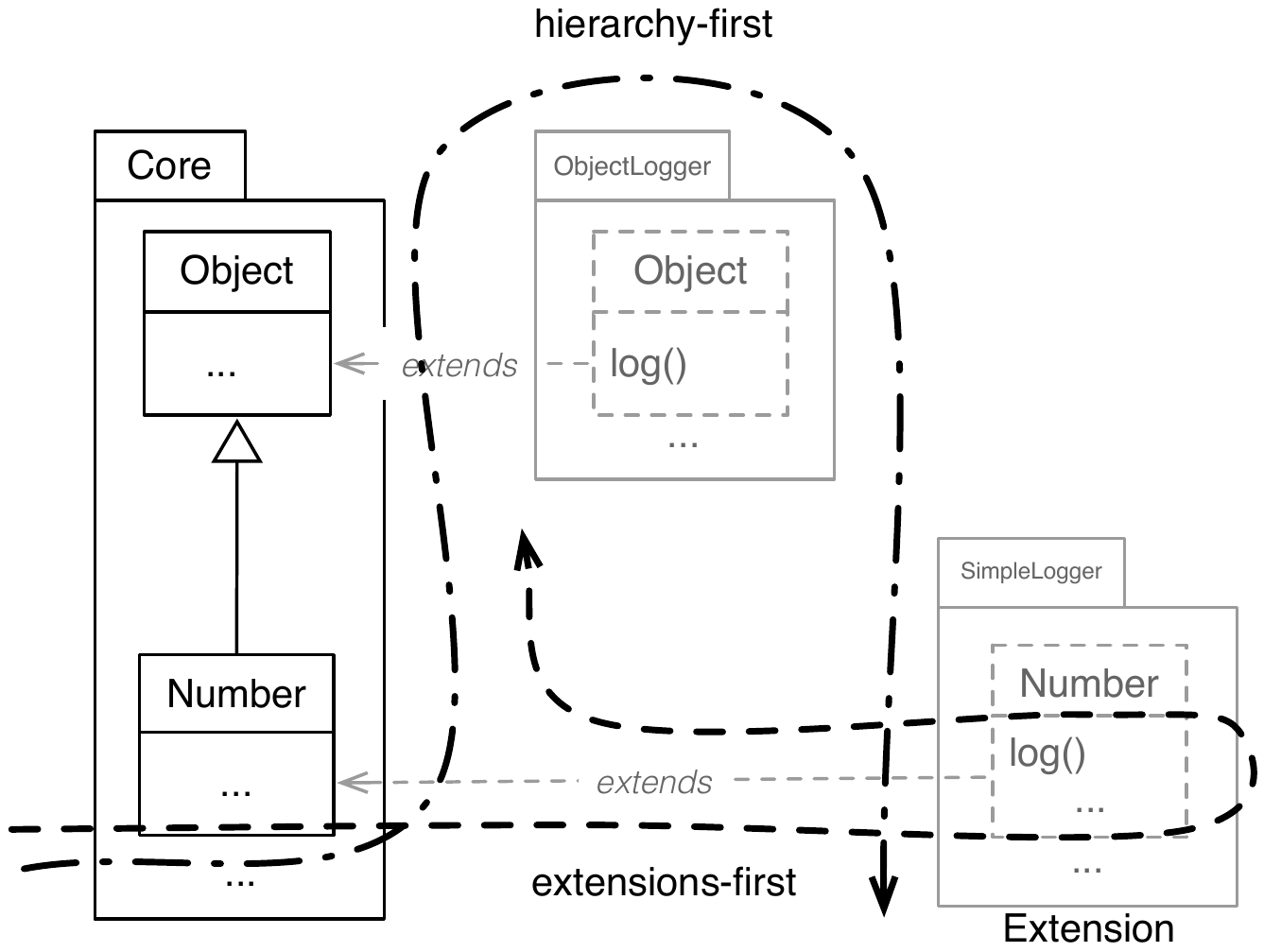}
    \caption{Two method selection strategies: extensions-first and hierarchy first.}
    \label{fig:selection}
    \end{figure}

Once the sequence of active extensions are determined according to one of the previous strategies, the second step is to select one method from all these extensions.
One strategy is to lookup a method in the first active extension throughout the hierarchy and then continue with following extensions (See \autoref{fig:selection}).
We refer to this strategy as \emph{hierarchy-first method selection strategy}.
Another solution is to lookup the method in the receiver class for each active scope in order and then continue to the superclass.
We refer to this strategy as \emph{extensions-first selection strategy}.

The choice of the method selection strategy has a big impact for the accidental override depicted in \autoref{fig:override}.
Indeed, given a sequence of active extensions, these strategies determine whether in a hierarchy two extension methods with the same name from different extensions have an override relationship or not.

\paragraph{Extensions-first.} It searches for a suitable method in each active extension before searching in the superclass of the receiver class.
This is the strategy used by all solutions presented in \autoref{sec:solutions}.
\[
  \fn{select_{ext}}(c, s, \exts) =
  \begin{cases}
    \fn{lookupInClass}(c,s,\exts)					& \text{if it is defined}\\
    \fn{select_{ext}}(\fn{superclass}(c),s,\exts) 		& \text{else if $\fn{superclass}(c)$ is defined} \\
    \text{is undefined}							& \text{otherwise} \\
  \end{cases}
\]

The function $\fn{select_{ext}}$ first looks if the first of extension in $\exts$ defines a method for the provided class and signature using the function $\fn{lookupInClass}$.
If no method is found (\ie $\fn{lookupInClass}(c,s,\exts)$ is undefined), $\fn{select_{ext}}$ continues recursively with the superclass of $c$ if it exists.
Otherwise, it is undefined if $\fn{superclass}(c)$ is undefined.
The function $\fn{lookupInClass}$ searches for the first suitable method defined for a given class in a given sequence of extensions.
It is defined as follow:
\[
\fn{lookupInClass}(c, s, <>)  \text{ is undefined}
\]
\[
\fn{lookupInClass}(c, s, \exts) =
\begin{cases}
  \fn{method}(c,s,\exts_1)				& \text{if it is defined}\\
  \fn{lookupInClass}(c,s, \fn{tail}(\exts) )	& \text{otherwise}\\
\end{cases}
\]

With extension-first method selection, a method can be overridden in extensions with higher priority in the class of the method or in any active extensions in subclasses of that method class.

\paragraph{Hierarchy-first.} It first looks up for the whole hierarchy of the receiver class in the context of the first extension and then consider the other extensions.
As we will see later, this solution permits to limit occurrence of accidental overrides.
\[
  \fn{select_{hrc}}(c,s,<>) \text{ is undefined}
\]
\[
\fn{select_{hrc}}(c,s,\exts) =
\begin{cases}
	\fn{lookupInExtension}(c,s, \exts_1)	& \text{if it is defined}\\
	\fn{select_{hrc}}(c,s,\fn{tail}(\exts))	& \text{otherwise}
\end{cases}
\]

The function $\fn{select_{hrc}}$ first looks if the first scope of $\exts$ defines a method in the provided $c$ or in class $c$ inherits from thanks to the function $\fn{lookupInExtension}$.
If no method is found, it continues recursively with the remaining scopes if there is some.
The function $\fn{lookupInExtension}$ searches for the first method defined in the hierarchy of a given class in a given extension.
It is defined as follow:

\[
  \fn{lookupInExtension}(c, s, \ext) = \\
  \begin{cases}
    \fn{method}(c,s,\ext)						& \text{if it is defined} \\
    \fn{lookupInExtension}& \text{else if $\fn{superclass}(c)$ is defined} \\
    \quad(\fn{superclass}(c),s,\exts) 	\\
    \text{is undefined}							& \text{otherwise} \\
  \end{cases}
\]

With hierarchy-first method selection, an extension method imported in a subclass can be overridden by extension methods imported by superclasses, if the extensions in the superclasses have higher priority.

\section{Comparison and Discussion}
\label{sec:analysis}

This section extends the comparison criterion with import granularities.
Then, we provide a comparison of existing solutions to expose the concepts.
We include in this comparison solutions for statically-typed languages to show how our conceptual decomposition captures also their semantics.
We present an analysis of the studied approaches.

\subsection{Declaration of Dependencies}
\label{sec:dependencies}
Once extension methods are local to their users it is mandatory for the users to declare which extension methods they bring into scope.
Hence, all the existing solutions here solve the problem of hidden dependencies.
These dependencies are usually declared with some form of import statements.
Such an import statement between a user (the \emph{importer}) and a set of extension methods (the \emph{importee}) requires answering two questions:
``What is imported?" and ``Where is it imported?".

\paragraph{Importee granularity.}
How can a developer declare which extension methods should be imported?
Many different granularities can be considered.
Importing extension methods one by one is tedious: the solutions presented here offer means to group related extension methods together.
One possible grouping is at the class-extension level (used by Classboxes for example) \ie extension methods are grouped by the class they extend.
This kind of grouping is simple but cannot specify a set of related methods in different classes (such as the \ct{asParser} methods presented in \autoref{sec:petitparser}).
Also, with classboxes and method shelters this class-centric grouping cannot specify different sets of methods for the same class.
Being able to make different groups for the same class can be useful: one group for a public API while another group is not meant to be exposed because it is for implementation purpose only.
Another possible grouping is the extension group (used by Method Shelters, Refinements and Categories) where extension methods are grouped under a named extension and can affect different classes.

\paragraph{Importer granularity.}
To which extent/scope is visible an extension?
With Classboxes for example, a class extension is imported and visible for all methods in the importing Classbox.
With Groovy Categories, extension methods are active during the execution of an importing block.
With Ruby Refinements, imported extension methods are visible in the importing class and all its subclasses.

\subsection{Comparison of Solutions}

We summarize in \autoref{fig:comparison} a comparison of existing scoped extension methods solutions according to the following criteria previously discussed: the importee granularity, the importer granularity, the extension activation strategy and the method selection strategy.

We observe in the table that solutions for dynamically-typed languages (Matriona and Classboxes on Squeak, Method Shelters and Categories on Groovy) use mainly local rebinding solutions.
On the other hand, solutions for statically-typed languages (MultiJava and Expanders on Java, PRM Refinements) use lexical activations.
The one exception is Ruby Refinements that uses lexical activations on a dynamically-typed language.

\begin{table}[hbt]
\centering
\caption{Comparison of the different approaches to scoped extension methods}
\footnotesize
\begin{tabular}{lcccc}
\toprule
        				& Importee 		 		& Importer					& Extension  				& Method  \\
				& granularity				& granularity				& activation strategy			& selection strategy \\
        \midrule
        Classboxes 	& one class extension 		& package				& bottom-up 				& extensions-first \\
        \cite{shortBerg03a,shortBerg05b,Berg05c}				& per class  		&						& local rebinding 			& \\
                &per package\\
                  \cmidrule(l){2-5}
        Method  		& one extension          		& package 				& controlled 		 		& extensions-first \\
        Shelters~\cite{Akai12a}		& per package				& 						& bottom-up				& \\
        				&						&						& local rebinding			& \\
                  \cmidrule(l){2-5}
        Groovy	& many extensions 			& block of code 			& top-down  				& extensions-first \\
        Categories~\cite{Koen07a}  & per package				& 						& local rebinding			& \\
                  \cmidrule(l){2-5}
                  Matriona~\cite{Spri16a} 	& many extensions & package & controlled & extensions-first \\
                  & per package	&		& top-down	& \\
                  &	& & local rebinding			& \\
                    \cmidrule(l){2-5}
                  Ruby 	& many extensions			& class and  				& lexical 					& extensions-first \\
       	Refinements			& per package		 		& its subclasses			&						& \\
              \cmidrule(l){2-5}
        MultiJava~\cite{shortClif00a} &	many extensions & package & lexical	& hierarchy-first \\
        & per package\\
            \cmidrule(l){2-5}
        Expanders~\cite{shortWart06a} 	& many extensions & package & lexical & hierarchy-first \\
                & per package\\
            \cmidrule(l){2-5}
        PRM~\cite{Duco07a} 	& many extensions & package & lexical & hierarchy-first \\
          & per package\\
      \bottomrule
\end{tabular}
\label{fig:comparison}
\end{table}
\subsection{Discussion}

As one of the authors of local-rebinding~\cite{shortBerg03a,shortBerg05b}, and after several years gathering more experience on the topic, we believe that local-rebinding brings more problems than solutions.
Accidental contextual overriding asides, local-rebinding violates object encapsulation since the same object can behave differently depending on the caller's code.
Lexical activation of extension does not have this problem. Indeed, we consider that lexically-activated extension methods do not cause accidental overrides but just normal intentional overrides because developers know beforehand which extensions are active in a scope and how they may override each other. Therefore they have a simpler and more predictable behavior that allows for local reasoning of a program.

The design space of scoped method extensions is wider than one can expect.
For instance, the active extension strategy is not the only design choice.
A language designer should also thing about method-selection, import granularity, security and so on.
For example, we determined that while local-rebinding improves code adaptability it causes too much encapsulation problems.
Despite lexical extension methods cannot modify the behavior of an object in a contextual manner like local-rebinding, they are easier to reason about.

The import relationship granularity has consequences on expressivity and segregation of extension methods in meaningful groups.
From the importee perspective, we saw that being able to define extensions \ie named groups of extension methods is the best solution.
Extension groups are more powerful than class extensions because:

\begin{itemize}
\item an extension can specify a set of related methods in different classes (such as the \ct{asParser} methods presented in \autoref{sec:petitparser}),
\item different extensions can specify different sets of methods for the same class,
\item and the previous class-based grouping can be realized with extensions whose methods all belong to the same class.
\end{itemize}

Methods, classes and packages are all valuable importers granularities and a solution should support all of them.
However, this requires the language to support grouping of methods.
The imports taken into account for a regular method would be the imports declared at the method-level plus the imports declared at the class-level, plus the imports declared at the package-level.
This includes the associated trade-off of increasing the language complexity.

Finally, whereas we called some overrides as ``accidental", they can also be malicious, \eg voluntarily corrupting a class behavior to gain access to protected operations or break fundamental invariants.
This is why the design of scoped extension methods and method selection strategy are crucial because they have a strong impact on accidental overrides as we will see in the next section.

\section{Our Formal Framework In Action: Example of an Analysis to Minimize Accidental Overrides}
\label{sec:minimizingAOS}

As discussed in previous sections, extension methods are useful but accidental overrides are insidious, hard to detect and may be frequent when several packages are involved in a program.
For example, in Pharo 3\footnote{build \#860, which contains 4115 classes/traits and 74648 methods}, 4.7\% of all methods are extension methods, 16.7\% of all classes and traits are extended, 48.1\% of all packages define an extension method and 31.7\% of all packages define a class or a trait that is extended by another package.
These numbers illustrate that in such a practical context, extension methods pose a high risk of accidental overrides limited thanks to coding conventions.

In this section we show how we can use our defined formal framework to propose a metric to estimate the risk of accidental overrides for the two method selection strategies: extension-first and hierarchy-first.
We define our metric and use it to determine which strategy provides the least risk.
The objective of this section is not to provide bullet-proof metric but to show how our formal framework can be used for this purpose.
Language designers are free to not follow this metric.

\paragraph{Accidental Overriding Space (AOS).} We call accidental override space (AOS) the set of all possible locations where a method could accidentally override another method for a given message.
For example, let us consider an arbitrary message $mess = (c_{rcv}, s, \exts)$ with signature $s$ sent to an instance of $c_{rcv}$ with the sequence of active extensions $\exts$.
Let $meth = \fn{method}(c_{def},s,e_i)$ be the method this message dispatches to.
This method is declared in the extension $\ext_i$, the \emph{i}-th extension of $\exts$ (possibly $\globalExtension$ if it is a regular method) for the class $c_{def}$ (\ie $c_{rcv}$ or one of its superclasses).
Now let us consider the addition of an arbitrary method $new = \fn{method}(c_{new},s,e_j)$ with the same signature $s$ in extension $e_j$.
We want to model the set of method locations where this new method would cause an accidental override, \ie the set of method locations that would cause $mess$ to dispatch to $new$ instead of $meth$.
Since the method $new$ has the same signature as $meth$, a method location only consists of a class and an extension.
If $new$ overrides $meth$ and are defined in the same extension $\ext_i$, this override is intentional, so we only consider locations where $j \neq i$.

\paragraph{AOS of Extension-First Strategy ($\fn{AOS_{ext}}$).}
For extension-first method selection, $new$ accidentally overrides $meth$ if:
(1) $new$ is defined for a subclass of $c_{def}$ in an extension $e_j$ in $\exts$ where $j \neq i$, or
(2) $new$ is defined for $c_{def}$ in an extension $e_j$ where $j < i$.
If we note $\superclassInv(c)$ all the subclasses of a class $c$ (transitive closure of the inverse of $\fn{superclass}$), we have:
\vspace{-1ex}
\[
\begin{array}{l}
\fn{AOS_{ext}}(mess, meth) =
\{ (c_{new}, s, e_j)  ~|~\\
\hspace{3cm}(c_{new} \in \superclassInv(c_{def})  \wedge i \neq j) \vee (c_{new} = c_{def} \wedge i < j)
\}
\end{array}
\]

\noindent
The size of $\fn{AOS_{ext}}(mess, meth)$ is given by:
\[|\fn{AOS_{ext}}(mess, meth)| = | \superclassInv(c_{def}) | \times (| \exts | - 1) + (i - 1)\]

\paragraph{AOS of Hierarchy-First Strategy ($\fn{AOS_{hrc}}$).}
For hierarchy-first method selection, $new$ accidentally overrides $meth$ if $new$ is defined for any class in $c_{def}$ hierarchy in an extension $e_j$ in $\exts$ where $j < i$.
If we note $\superclasses(c)$ all the superclasses of a class $c$, we have:
\[
   \begin{array}{l}
\fn{AOS_{hrc}}(mess, meth) = \{ (c_{new}, s, e_j)  ~|~ \\
\hspace{3cm}c_{new} \in (\superclassInv(c_{def}) \cup c_{def} \cup \superclasses(c_{def})) \wedge i < j \}
   \end{array}
\]

\noindent
The size of $\fn{AOS_{hrc, meth}}(mess)$ is given by:
\[|\fn{AOS_{hrc}}(mess, meth)| = (|\superclassInv(c_{def})| + |\superclasses(c_{def})| + 1) \times (i-1) \]

\paragraph{Comparison of $|\fn{AOS_{ext}}|$ and $|\fn{AOS_{hrc}}|$.}
We can now compare the AOS of each method selection strategy.
We ask ourself when the hierarchy-first strategy is better than the extension-first strategy \ie when $|\fn{AOS_{hrc}}(mess, meth)| \leq |\fn{AOS_{ext}}(mess, meth)|$.

\paragraph{AOS Comparison in the Pharo Language.}
To compare the two metrics defined above, we must have an idea of the average number of subclasses and superclasses a class has.
This section shows how this metric is concretized in the case of Pharo. A more general language-independent approach follows after this analysis.

In Pharo the average number of subclasses of a class is $8.82$ and the average number of superclasses of a class is $3.83$.
With these numbers our inequality reduces to $1.43 i - 0.43  \leq |\exts|$.
In the table below, the first row shows $|\exts|$ ranging from $1$ to $10$.
Remember that $i$ can range from $1$ to $|\exts|$.
For each value of $|\exts|$, the second row shows the maximum value of $i$ that still satisfies the inequation above.
This table shows that hierarchy-first strategy (second row) has less risk to cause accidental overrides than extension-first strategy (first row).

\vspace{1ex}
\begin{center}
 \begin{tabular}{c|cccccccccc}
      	\toprule
        $| \exts |$ & 1 & 2 & 3 & 4 & 5 & 6 & 7 & 8 & 9 & 10\\
        $ i \leq $ & 1 & 1 & 2 & 3 & 3 & 4 & 5 & 5 & 6 & 7 \\
        \bottomrule
\end{tabular}
\end{center}
\vspace{1ex}

For example, when $5$ extensions are active ($|\exts|=5$), the hierarchy-first strategy has less risk to cause an accidental override than extension-first strategy whenever $meth$ belongs to one of the first $3$ active extensions ($i<=3$).
Following the table, we observe that for this example in 67\% of total cases hierarchy-first strategy has less accidental method override risks. .
Extension-first strategy performs better when $meth$ belongs to the last extensions \ie the ones that have smaller priority.
But it also means that using extension-first strategy, accidental overrides can happen for extension with a lower priority.
So, hierarchy-first strategy has also the advantage that an accidental method override can only happen for extension with a higher priority.
All of these reasons show that the hierarchy-first strategy is better to limit accidental overrides.

\paragraph{Generalization.} The above analysis relies on average number of subclasses and superclasses of a given class.
The following question then arises: Does hierarchy-first strategy always provide a smaller risk of accidental overrides than extension-first strategy?
If we make these two numbers varying from 1 to 10 and compute all results, it appears that: hierarchy-first performs better when the average number of
subclasses is greater than the average number of superclasses (and extension-first on the opposite way).
And this assertion is usually true because object-oriented hierarchies are built by extension \ie subclassing.

\paragraph{Conclusion.} This analysis shows that generally, the hierarchy-first strategy minimizes the risk of ``accidental" overrides in comparison the extension-first strategy.

\section{Related Work}
\label{sec:relatedwork}

We have already shown and analyzed in previous sections existing solutions for scoping extension methods. In this section, we compare this work with respect to the problem of conflicts and other module related formalizations.

\paragraph{Module Taxonomy.}
Bergel, Ducasse and Nierstrasz present a module taxonomy in their work ``Analyzing Module Diversity''~\cite{Berg05c}. They present a simple module calculus consisting of a small set of operators over environments and modules. Using these operators, they are then able to specify a set of module combinators that capture the semantics
of Java packages, C\# namespaces, Ruby modules, selector namespaces, gbeta classes, classboxes, MZScheme units, and MixJuice modules. The article develops a simple taxonomy
of module systems. Even if the paper covers Classboxes and selector namespaces, it does not specifically focus on method extensions and does not cover some of the more recent languages supporting class extensions such as Groovy and Method Shelters. In addition, their semantics does not capture the fine grained aspects of the local rebinding lookup stack traversal.

\paragraph{Accidental overrides.} Simple changes can have unexpected effects due to implicit contracts between a class and its subclasses. This well-known problem, coined as the \emph{fragile base class problem} \cite{shortMikh98a}, is due to the fact that current languages do not support well extension contracts: Just changing the calling structure of a method without changing its external behavior may have unexpected effects in presence of subclasses. C\# is the one of the rare languages that offers a way to control unintended name capture (called accidental overrides in this paper). C\# allows the programmer to qualify a method with the keyword \ct{new} (rather than \ct{override}) to declare that while the newly defined method has the same name as the one in its superclasses, it is used for a different concept than in the superclasses. As such, all calls in the superclass hierarchy that would invoke a method with the same name will not consider the new method.

%

\section{Conclusion}
Globally-visible extension methods can lead to conflicts: accidental overrides and overwrites.
These conflicts pose class encapsulation problems that can lead to subtle bugs or be exploited by malicious parties.
In this article, we analyzed multiple solutions that propose to scope extension methods in dynamically-typed languages: Classboxes, Ruby Refinements, Method Shelters, and Groovy Categories.

We defined scoped extension mechanisms as a combination of a \emph{active extension strategy} and a \emph{method selection strategy}. An active extension strategy defines what extension methods are available in a given context. We identified lexical activation as well as two flavours of local-rebinding activations that were partially described in the literature.
A method selection strategy defines how a method is selected when there are multiple active extensions defining methods with the same signature.
Method selection strategies can give precedence to the class hierarchy (hierarchy-first) or to the extensions (extensions-first).

We then used these formal semantics to characterize other solutions such as MultiJava, Expanders and Matriona.
We show that the semantics of scoped extension methods has a big impact on accidental overrides, and concluded that the combination of lexical extension methods with the hierarchy-first method selection strategy gives the best results to minimize them.

\acks We thank the french DGA (Direction G\'en\'erale de l'Armement) for the PhD grant of Camille Teruel.
\printbibliography


\end{document}